\documentstyle[12pt,psfig,epsfig,axodraw]{article}
\advance\textheight by \topskip
\newcommand{\newc}{\newcommand}

\def\Ord{\lower .7ex\hbox{$\;\stackrel{\textstyle <}{\sim}\;$}}
\def\OOrd{\lower .7ex\hbox{$\;\stackrel{\textstyle >}{\sim}\;$}}

\newc{\anti}{\overline}
\newc{\be}{\begin{equation}}
\newc{\ee}{\end{equation}}
\newc{\br}{\begin{eqnarray}}
\newc{\er}{\end{eqnarray}}
\newc{\ba}{\begin{array}}
\newc{\ea}{\end{array}}
\newc{\bi}{\begin{itemize}}
\newc{\ei}{\end{itemize}}
\newc{\bn}{\begin{enumerate}}
\newc{\en}{\end{enumerate}}
\newc{\bc}{\begin{center}}
\newc{\ec}{\end{center}}
\newc{\ul}{\underline}
\newc{\ol}{\overline}
\newc{\ra}{\rightarrow}
\newc{\lra}{\longrightarrow}
\newc{\wt}{\widetilde}
\newc{\ti}{\times}
\newc{\Dir}{\kern -6.4pt\Big{/}}
\newc{\Dirin}{\kern -10.4pt\Big{/}\kern 4.4pt}
\newc{\DDir}{\kern -10.6pt\Big{/}}
\newc{\DGir}{\kern -6.0pt\Big{/}}
\newc{\lam}{\lambda}
\newc{\gam}{\gamma}
\newc{\eps}{\epsilon}
\newc{\kap}{\kappa}
\newc{\modulus}[1]{\left| #1 \right|}
\newc{\eq}[1]{(\ref{eq:#1})}
\newc{\eqs}[2]{(\ref{eq:#1},\ref{eq:#2})}
\newc{\etal}{{\it et al.}\ }
\newc{\ibid}{{\it ibid}.}
\newc{\eg}{{\it e.g.}\ }
\newc{\ie}{{\it i.e.}\ }

\newc{\nonum}{\nonumber}
\newc{\lab}[1]{\label{eq:#1}}
\newc{\dpr}[2]{({#1}\cdot{#2})}
\newc{\lsimeq}{\stackrel{<}{\sim}}
\newc{\lt}{\stackrel{<}}
\newc{\gt}{\stackrel{>}}
\newc{\gsimeq}{\stackrel{>}{\sim}}

\newc{\half}{\frac{1}{2}}

\def\lapp{\mathrel{\rlap{\raise.5ex\hbox{$<$}}
                    {\lower.5ex\hbox{$\sim$}}}}
\def\gapp{\mathrel{\rlap{\raise.5ex\hbox{$>$}}
                    {\lower.5ex\hbox{$\sim$}}}}
\newc{\bQ}{\overline{Q}}
\newc{\dota}{\dot{\alpha }}
\newc{\dotb}{\dot{\beta }}
\newc{\dotd}{\dot{\delta }}
\newc{\nindnt}{\noindent}
\newc{\matth}{\mathsurround=0pt}
\def\ML{\ifmmode{{\mathaccent"7E M}_L}
             \else{${\mathaccent"7E M}_L$}\fi}
\def\MR{\ifmmode{{\mathaccent"7E M}_R}
             \else{${\mathaccent"7E M}_R$}\fi}

\newc{\thetaw}{\theta_W}
\newc{\sm}{${\cal {SM}}$}
\newc{\as}{\alpha_s}
\newc{\aem}{\alpha_{em}}

\newc{\ppbar}{\mbox{$p\overline{p}$}}
\newc{\bbbar}{\mbox{$b\overline{b}$}}
\newc{\ccbar}{\mbox{$c\overline{c}$}}
\newc{\ttbar}{\mbox{$t\overline{t}$}}
\newc{\eebar}{\mbox{$e\overline{e}$}}
\newc{\zzero}{\mbox{$Z^0$}}

\newc{\wplus}{\mbox{$W^+$}}
\newc{\wminus}{\mbox{$W^-$}}
\newc{\elp}{\mbox{$e^+$}}
\newc{\elm}{\mbox{$e^-$}}
\newc{\elpm}{\mbox{$e^{\pm}$}}
\newc{\qbar}     {\mbox{$\overline{q}$}}
\newc{\epem}{\mbox{$e^+e^-$}}
\newc{\lp}{\mbox{$e^+$}}
\newc{\lm}{\mbox{$e^-$}}
\newc{\lpm}{\mbox{$e^{\pm}$}}
\newc{\lplm}{\mbox{$\ell^+\ell^-$}}

\newc{\susy}{{{SUSY}}}

\def\photino{\ifmmode{\mathaccent"7E \gamma}\else{$\mathaccent"7E \gamma$}\fi}
\def\taugluino{\ifmmode{\tau_{\mathaccent"7E g}}
             \else{$\tau_{\mathaccent"7E g}$}\fi}
\def\mphotino{\ifmmode{m_{\mathaccent"7E \gamma}}
             \else{$m_{\mathaccent"7E \gamma}$}\fi}
\newc{\gluino}   {\mbox{$\wt{g}$}}
\newc{\mgluino}  {\mbox{$M(\gluino)$}}
\newc{\gl}{\wt g}
\newc{\mgl}{m_{\gl}}
\def \lspone{\wt\chi_1^0}
\def \mlspone{m_{\lspone}}

\def \snu{\wt{\nu}}

\def \snul{\snu_{\ell}}

\def \slepl{\wt{l}_L}
\def \mslepl{m_{\slepl}}
\def \slepr{\wt{l}_R}
\def \mslepr{m_{\slepr}}

\def \ul{\wt{u}_L}

\def \sql{\wt{q}_L}
\def \msql{m_{\sql}}
\def \sqr{\wt{q}_R}
\def \msqr{m_{\sqr}}

\newc{\squark}   {\mbox{$\wt{q}$}}
\newc{\msquark}  {\mbox{$m(\squark)$}}
\newc{\msq}{m_{\squark}}
\newc{\sqbar}    {\mbox{$\bar{\wt{q}}$}}
\newc{\ssb}      {\mbox{$\squark\overline{\squark}$}}

\newc{\csquark}  {\mbox{$\wt{c}$}}
\newc{\csquarkl} {\mbox{$\wt{c}_L$}}
\newc{\mcsl}     {\mbox{$m(\csquarkl)$}}

\def \lstop{\wt{t}_{1}}

\def \mlstop{m_{\lstop}}

\def \lstoplstop{\lstop\lstop^*}

\newc{\tsquark}  {\mbox{$\wt{t}$}}
\newc{\ttb}      {\mbox{$\tsquark\overline{\tsquark}$}}
\newc{\ttbone}   {\mbox{$\tsquark_1\overline{\tsquark}_1$}}
\newc{\mix}{\theta_{\wt t}}
\newc{\cost}{\cos{\theta_{\wt t}}}
\newc{\costloop}{\cos{\theta_{\wt t_{loop}}}}
\newc{\tb}{\tan\beta}
\newc{\vev}[1]{{\left\langle #1\right\rangle}}

\def \abot{A_{b}}

\def \atau{A_{\tau}}
\newc{\mhalf}{m_{1/2}}
\newc{\mgut}{M_U}
\newc{\mzero} {\mbox{$m_0$}}

\newc{\lampp}{\lam^{\prime\prime}}
\newc{\lamp}{\lam^{\prime}}
\newc{\rpv}{\not R_p}
\newc{\rp}{R_{p}}
\newc{\lamotho}{\lam_{131}}
\newc{\lampotho}{\lam'_{131}}
\newc{\lamdpotho}{\lam''_{131}}

\def \br{\emph{BR}}
\def \branche{\br(\lstop\ra be^{+}\nu_e \lspone)\times \br(\lstop^{*}\ra \bar{b}q\bar{q^{\prime}}\lspone)}
\def \branchmu{\br(\lstop\ra b\mu^{+}\nu_{\mu} \lspone)\times \br(\lstop^{*}\ra \bar{b}q\bar{q^{\prime}}\lspone)}
\def\Ecm{\ifmmode{E_{\mathrm{cm}}}\else{$E_{\mathrm{cm}}$}\fi}
\newc{\rts}{\sqrt{s}}
\newc{\gev}{\,GeV}
\newc{\mev}{~{\rm MeV}}
\newc{\tev}  {\mbox{$\;{\rm TeV}$}}
\newc{\gevc} {\mbox{$\;{\rm GeV}/c$}}
\newc{\gevcc}{\mbox{$\;{\rm GeV}/c^2$}}
\newc{\intlum}{\mbox{${ \int {\cal L} \; dt}$}}
\newc{\call}{{\cal L}}
\def \miset{\not\!\!{E_T}}
\newc{\etmiss}{/ \hskip-7pt E_T}
\def \met  {\mbox{${E\!\!\!\!/_T}$}}
 
\newc{\ptmiss}{/ \hskip-7pt p_T}
\newc{\ifb}{\mbox{${\rm fb}^{-1}$}}
\newc{\ipb}{\mbox{${\rm pb}^{-1}$}}
\newc{\chis}{\mbox{$\chi^{2}$}}
\newc{\pt}{\mbox{$p_T$}}
\newc{\et}{\mbox{$E_T$}}
\newc{\dedx}{\mbox{${\rm d}E/{\rm d}x$}}
\newc{\pb}{~{\rm pb}}
\newc{\fb}{~{\rm fb}}
\newc{\ycut}{y_{\mathrm{cut}}}
\def \ppbar{p\bar{p}}

\def \epem{e^+e^-}

\def \d0{\emph{D0 }}

\def \jet(s){\emph{jet(s) }}

\def\loopdk{\lstop \ra c \lspone}
\def\brloopdk{\br(\loopdk)}

\newc{\mplanck}{M_{\rm P}}

\def\issue(#1,#2,#3){{\bf #1} (#3) #2 } % PLB format! Vol, (Year), page
\def\PRD(#1,#2,#3){Phys.\ Rev.\ D \issue(#1,#2,#3)}
\def\NPB(#1,#2,#3){Nucl.\ Phys.\ B \issue(#1,#2,#3)}
\def\JP(#1,#2,#3){J.\ Phys.\issue(#1,#2,#3)}
\def\PL(#1,#2,#3){Phys.\ Lett. \issue(#1,#2,#3)}
\def\PLB(#1,#2,#3){Phys.\ Lett.\ B  \issue(#1,#2,#3)}
\def\ZP(#1,#2,#3){Z.\ Phys. \issue(#1,#2,#3)}
\def\ZPC(#1,#2,#3){Z.\ Phys. \ C  \issue(#1,#2,#3)}
\def\PREP(#1,#2,#3){Phys.\ Rep. \issue(#1,#2,#3)}
\def\PRL(#1,#2,#3){Phys.\ Rev.\ Lett. \issue(#1,#2,#3)}
\def\MPL(#1,#2,#3){Mod.\ Phys.\ Lett. \issue(#1,#2,#3)}
\def\RMP(#1,#2,#3){Rev.\ Mod.\ Phys. \issue(#1,#2,#3)}
\def\SJNP(#1,#2,#3){Sov.\ J. \ Nucl.\ Phys. \issue(#1,#2,#3)}
\def\CPC(#1,#2,#3){Comp.\ Phys. \ Comm. \issue(#1,#2,#3)}
\def\IJMPA(#1,#2,#3){Int.\ J. \ Mod. \ Phys.\ A \issue(#1,#2,#3)}
\def\MPLA(#1,#2,#3){Mod.\ Phys.\ Lett.\ A \issue(#1,#2,#3)}
\def\PTP(#1,#2,#3){Prog.\ Theor.\ Phys. \issue(#1,#2,#3)}
\def\RMP(#1,#2,#3){Rev.\ Mod.\ Phys. \issue(#1,#2,#3)}
\def\NIMA(#1,#2,#3){Nucl.\ Instrum.\ Methods \ A \issue(#1,#2,#3)}
\def\JHEP(#1,#2,#3){J.\ High \ Energy \ Phys. \issue(#1,#2,#3)}
\def\EPJC(#1,#2,#3){Eur.\ Phys.\ J. \ C \issue(#1,#2,#3)}
\def\RPP (#1,#2,#3){Rept.\Prog.\Phys \issue(#1,#2,#3)}
\newc{\PRDR}[3]{{Phys. Rev. D} {\bf #1}, Rapid  Communications, #2 (#3)}
\begin{document}
\begin{titlepage}
%\begin{center}
%{\Large DRAFT V1.0}
%\end{center}
\begin{flushright}
\today\\
{hep-ph/0404049}
\end{flushright}
\vskip .6cm
\begin{center}
{\Large\bf  Top squark mass: current limits revisited and new limits 
from Tevatron Run-I
}\\[1.00cm]
{\large Siba Prasad Das$^{a,}$,\footnote{\it spdas@juphys.ernet.in}
Amitava Datta$^{a,}$} \footnote{\it adatta@juphys.ernet.in}
%\\[0.3 cm]
{and}
%\\[0.3 cm]
{\large Manas Maity$^{b,}$} \footnote{\it $iam\_manas\_maity$@yahoo.co.in }
{\\[0.3 cm]}
{\it $^a$ Department of Physics, Jadavpur University,
Kolkata 700032,India}\\[0.3cm]
{\it $^b$  Department of Physics, Visva-Bharati, Santiniketan 731235
India.}\\[0.3cm]

\end{center}
\vspace{.1cm}

\begin{abstract} 
{\noindent\normalsize Analyzing the $\ell +n$-$jets+
\met$ ( where $n \ge 2 $ ) data from Run-I of the Tevatron using the
Bayesian technique, we obtain model independent limits on the product
$\br(\lstop\ra be^{+}\nu_e \lspone) \times \br(\lstop^{*}\ra
\bar{b}q\bar{q^{\prime}}\lspone)$ for different values of the lighter top
squark ( $\lstop$) mass and the lightest supersymmetric particle (
$\lspone$) mass.  The  signal events have been simulated by interfacing the
4-body decay of $\lstop$ at the parton level with the event generator
PYTHIA.  These limits have been translated into exclusion plots in the
$\mlstop$-$\mlspone$ plane, which also turn out to be fairly model
independent for fixed values of the BR of the competing loop decay mode
$\lstop\ra c \lspone$. Assuming the loop decay BR to be negligible and
using the leading order cross section for $\lstoplstop$ pair production,
we obtain conservatively $\mlstop \ge $77.0 (74.5) GeV for
$\mlspone$=5(15) GeV, while for $\br(\lstop\ra c \lspone)$=20\%, the
corresponding limits are $\mlstop \ge $68.0 (65.0) GeV. Using the larger
next to leading order cross-section stronger limits are obtained. For
example, if $\br(\lstop\ra c \lspone)$=20\%, $\mlstop \ge $73.0 (72.7) GeV
for $\mlspone$=5(15) GeV. Our limits nicely complement the ALEPH bounds
which get weaker for low $\mlspone$.

}
\end{abstract}

\end{titlepage}

\textheight=8.9in

%%%
The Minimal Supersymmetric Standard Model (MSSM)\cite{susy} is a  
well motivated extension of the Standard Model(SM), but there  is
no evidence of it as well as it has not been ruled out by the 
electroweak precision measurements at LEP\cite{leppre}.
Unfortunately, we are not equipped with any theoretical guideline
about the range of superparticle masses since the exact SUSY breaking 
mechanism is unknown yet, although several interesting suggestions 
exist\cite{susy}. From unsuccessful  searches at LEP~\cite{lepbound}
and Tevatron  Run-I~\cite{xsusy,work} some experimental lower bounds 
on superparticle  masses exist.

The second phase of  experiments at the Tevatron,
the Run-II, is in progress. It is expected that  an integrated
luminosity of at least 2 fb$^{-1}$ per experiment at 2 TeV center of mass
energy will be accumulated. This is about ten times larger than the 
acquired luminosity in Run-I with center of mass energy 1.8 TeV. 

However, in view of the existing limits on the masses of the strongly
interacting sparticles (squarks and gluinos) \cite{xsusy,work} and the
rather marginal increase in the center of mass energy, most of the unexplored
parameter space in this sector is likely to be beyond the kinematic reach
of Run-II as well. Since this is the only currently available machine for
direct SUSY searches until the LHC starts, it is important to identify the
sparticles with reasonable production cross sections which may be within
the striking range of the Tevatron.
                                                                                
The lighter top squark mass eigenstate $\lstop$ could be 
an interesting possibility. This is because  the large top quark mass
induces a large mixing term in the top squark mass matrix~\cite{stopmix}. 
When the matrix is diagonalized, one of the mass eigenvalues may turn 
out to be  rather small over a large region of the MSSM parameter space.
In fact, it is quite conceivable that $\lstop$  is
the next lightest supersymmetric particle (NLSP), the lightest
neutralino $\wt\chi_1^0$  being  the lightest supersymmetric
particle (LSP) by assumption in most R-parity($\rp$) conserving  models.

Since  the $\lstop$ could be  the only  strongly interacting
sparticle within the kinematic range of Run-II, 
it is important to carefully plan  the  strategy for searching it. 
The existing limits on $\mlstop$ may provide important guidelines 
for this plan. In the first part of this letter we shall critically re-examine
the existing limits. Since we do not want to commit ourselves to any specific
SUSY breaking mechanism we shall discuss only the limits which are valid
in the most general $\rp$ conserving MSSM. In the second part of this 
paper we shall derive some new limits using Run-I data.

The collider signatures, however, crucially depend on whether the top
squark is the NLSP or not. In this letter we shall be mainly concerned with
the scenarios with a top squark NLSP with $\mlstop$ below the top quark mass. 
It is further assumed that all three body decays like
$\lstop\ra  bW\lspone$, where  $\lspone$ is the only superparticle 
in the final state, are kinematically forbidden.
In this case the only allowed decay modes in the $\rp$ conserving MSSM
are the following:

(i)The flavour changing  loop decay into a charm quark and the LSP,
$\lstop \ra c \lspone$, ~\cite{hikasa} ; (ii) The 4-body decay into 
a b quark, the LSP and two approximately massless fermions
, $\lstop \ra b \lspone f \bar f'$, 
where $f\bar f'= q\bar q'~or~ l \bar \nu_{l} $( 
$\ell= e,\mu$)~\cite{boehm}.
 
We note in passing that if $m_W + \mlspone \lapp \mlstop \lapp m_b +
m_W + \mlspone$, then the decay $\lstop\ra  q W\lspone$, where q  
= d or s, can occur in principle. Of course the BR of this mode 
could be suppressed by a mixing angle expected to be very small if
the quark and the squark mass matrices are aligned. The magnitude
of this parameter, however, is very much model dependent and the
possibility that this mode may compete with the decays (i) and (ii)
also having small widths, can not be apriorily ruled out. The 
resulting signal consisting of W + light hadrons + $\miset$ may be difficult
to detect, especially so if $\mlspone$ and consequently the $\miset$ is small.
To the best of our knowledge this signal has not been studied so far.
This decay mode, which could  be a test of alignment of the quark and squark
mass matrices, is not of particular interest for this paper since Run-I data
is sensitive to $\mlstop \lapp m_W$ only.    

Until very recently most of the limits on the top squark NLSP, derived from 
unsuccessful searches at LEP and Tevatron, were based  on the assumption
that the former decay occur with 100\% branching ratio (BR). Moreover these
limits have additional dependence on SUSY parameters in the following way.

At hadron colliders the leading order (LO) cross section for pair
production of top squarks depends on $\mlstop$ only since it is a
pure QCD process\cite{lostop}. The dependence on
 other SUSY parameters, {\it e.g.}, ~the gluino mass $\mgl$, the masses of
the other squarks, the mixing angle $\cost$ ( where, $\mix$ is the mixing 
angle in top squark sector ), etc., arise only through the
next to leading order (NLO) corrections,  which yield  somewhat larger cross
sections\cite{nlostop}. The efficiency of the kinematical cuts required
to isolate the top squark signal from the SM background, on the other hand,
strongly depends on the lightest neutralino mass $\mlspone$. The existing
conservative limits from Tevatron based on the LO cross section 
\cite{cdfbound,tevbound} and the
 assumption of 100\% BR's of the loop decay, are presented as exclusion
plots in the $\mlstop$-$\mlspone$ plane (see Fig.(2) of \cite{cdfbound}). 
The most stringent bound, from Tevatron experiments,  puts a lower limit of
$m_{\wt t_1} \ge $ 119 GeV for $m_{\wt\chi^0_1}=$40 GeV.  This limit becomes 
considerably weaker for higher value of $m_{\wt\chi^0_1}$, e.g,
$m_{\wt t_1} \ge $ 102 GeV for
$m_{\wt\chi^0_1}=$50 GeV~\cite{cdfbound}. Thus, even if we temporarily set
aside the
questionable assumption of 100\% BR's for the loop decay, the existing limits
from Tevatron on $\mlstop$ could be rather weak 
for relatively large  $\lspone$ mass.

Using the model dependent assumption of the complete dominance
of the loop decay limits on $\mlstop$ have also been obtained at 
LEP\cite{lepboundstop}. At $e^+e^-$ colliders the electroweak $\lstop\lstop^*$
production cross section has an  additional dependence on the $\mix$.
The cross section is maximum for $\mix$ = 0$^{\circ}$ while it is minimum for 
$\mix$ = 56$^{\circ}$, when $\lstop$ decouples from the Z . For larger 
values of $\mix$ the cross section is essentially the same as that for 
$\mix$ = 56$^{\circ}$ \cite{bartl}, particularly so for relatively high 
$\mlstop$ kinematically  accessible to LEP . This behavior of the cross 
section ensures that the  limits corresponding to $\mix$ = 56$^{\circ}$ are 
valid to a very good approximation for higher values of $\mix$. The 
efficiency of the  kinematical cuts also depends on $\mlspone$ although 
the dependence is somewhat different from  that in the case of Tevatron 
data. For $\mix \gapp$ 56$^{\circ}$ and $\mlstop \gapp$ 78.0 GeV,  
no exclusion is possible for  low $\mlspone$, although for higher $\mlspone$ 
better limits are obtained even if $\mlspone \approx \mlstop$ (see Fig.2(a) of
~\cite{alephstop}). It should be emphasized that it is precisely
for these low $\mlspone$ the CDF limits using the same assumption  of the 
dominance of the loop decay are more  
stringent and limits extending beyond the kinematical reach  of LEP are
obtained. Thus the limits from LEP and Tevatron  complement each other. 

It has been known for some time that in a wide region of the MSSM
parameter space the BR's of the 4-body decay can be substantial and may even
dominate over the loop decay. The dependence of the 4-body decay rate on
SUSY parameters has been studied in great detail both in the MSSM
and  the minimal Supergravity (mSUGRA) \cite{boehm,admgspd}. Especially for 
large values of $\mix$ and small values of $\tb$ this mode can be the 
dominant one. Thus the limits discussed above may require significant revision. 
The dependence on other MSSM parameters will be reviewed later and some 
new points will be discussed.

Very  recently both  ALEPH \cite{alephstop} and D0 collaborations
\cite{d0stop}  have analyzed respectively LEP and Tevatron data using 
special assumptions for the 4-body decay.
D0 has obtained cross section  limits 
as a function of $\mlstop$ assuming 100\% BR's
for the leptonic 4-body decay\cite{d0stop}. This assumption, however,  
is unrealistic. As has already been noted this BR's does not  
exceed the 20\% level considering both the $e~ and ~\mu  $ 
modes \cite{admgspd}.

For the first time the ALEPH collaboration has analyzed the data
taking into account both the competing decay modes. One set of 
realistic limits have been obtained by assuming that the 4-body decay
overwhelms the loop decay and the relative BR's of the 4-body leptonic 
and hadronic decays of $\lstop$
closely follow that of the $W^*$ ( see Fig.3(a) of \cite{alephstop} ).
The exclusion plot in the  $\mlstop$-$\mlspone$ plane shows that
for low $\mlspone$, $\mlstop >$ 78.0 (84.0) GeV is allowed for 
$\mix $= 56$^{\circ}$ (0$^{\circ}$). 

Another set of limits has been obtained by varying both the loop decay 
and the 4-body semileptonic decay BR's as free parameters. 
They have then checked whether a particular $\mlstop$
can be excluded via any one of the two competing decay modes. As already
discussed, these limits also depend on $\mlspone$ and $\mix$. Unfortunately 
the numerical values of the semileptonic 4-body decay BR's used in deriving 
the limits  are not always realistic. For
example the absolute lower limit of $\mlstop > $ 63.0 GeV at 95\% confidence
level has been obtained for the loop decay BR's = 22\%, the semileptonic 4-body 
BR's = 55\%, $\Delta M = \mlstop - \mlspone =$ 5 GeV and $\mix = $ 56$^{\circ}$.
The assumed semileptonic 4-body BR's, however, is unrealistic. We have
checked that for $\mlstop$ within the kinematic reach of LEP the hadronic 
4-body BR's  is much larger than the semileptonic one over the entire
 MSSM parameter space.  Higher values of $\mlstop$ are excluded for lower 
values of $\mlspone$,  ( see Fig.4(a) and 4(b) of Ref.\cite{alephstop}). 
For example fixing the loop decay BR's at 22\% the strongest limit 
$\mlstop \gapp$ 95.0 GeV is obtained
for $\Delta M=$ 25 GeV. However for still larger values of $\Delta M$,
the limits get weaker as can be seen from the limit
$\mlstop \gapp$ 89.0 GeV  for 
$\Delta M=$ 45 GeV. No limit for still higher values of $\Delta M$  has
been presented. 
In summary  the ALEPH  limits become weaker for
 $\mix \gapp$ 56$^{\circ}$ and  relatively low $\mlspone$. 

The purpose of this letter is to show that precisely in these
regions of the MSSM parameter space, the data from Tevatron Run-I
\cite{cdfstop} already gives almost model independent,  
stronger limits inspite of the rather modest integrated luminosity.
The conservative limits using the LO production cross section can be
further improved if somewhat larger NLO cross sections are employed.
For most of the parameter space 
$\sigma_{NLO}\left(\ppbar\ra\lstop\lstop^*\right)$
is 30\% higher than the $\sigma_{LO}\left(\ppbar\ra\lstop\lstop^*\right)$
\cite{nlostop}. More importantly, this analysis outlines a strategy 
using which the Run-II search at slightly higher production cross section
and much higher integrated luminosity can spectacularly enrich  the
information about the top squark NLSP in a fairly model independent way.

%%%

We looked into the existing CDF and D0 data and tried to identify the 
one which can be best utilized to constrain the 4-body decay modes of
$\lstop$. In principle the classic $jets+$missing $E_T$ ($\met$) data
\cite{xsusy} used for 
squark-gluino searches  can be used to constrain the 4-body 
decay of the top squark in the all hadronic 
mode which has the largest BR's. Unfortunately the stiff $\met$ cut
used to extract the existing data suitable for heavy superparticle searches
give rather poor selection efficiency for the light $\lstop$ accessible 
at Tevatron Run-I. We, therefore, analyzed the CDF data\cite{cdfstop} 
used for a different search channel, $\lstop \ra b \ell^{+} \snul$,
leading to the signal $\ell +n$-$jets+\met$, where $n \ge 2$.
The same signal also arises
from the 4-body decay of top squark when one  $\lstop$ decays leptonically 
and the other decays hadronically. Our analysis, however, will be 
handicapped due to the rather modest branching ratio of the semileptonic  mode 
and the kinematical cuts optimized for a different decay channel.
Nevertheless some useful conclusions emerge. 

The 4-body decay has been simulated at the  parton level  using CTEQ4M
parton distribution function \cite{cteq}, where one
of the $\lstop$  decays leptonically while the other decays
hadronically{\footnote{Charge conjugate interactions have been
assumed unless otherwise stated.}: $$\ppbar\lra\lstoplstop\lra
b\bar{\ell}\nu_{\ell}\lspone\bar{b}q\bar{q^{\prime}} \lspone 
~~~( ~\ell~ =~ e ~or~ \mu~ ) $$
{\textsc PYTHIA} is then used for hadronization of the partons from
$\lstop$ decays. The final state particles have been passed through
a toy detector simulation ( using tools in {\textsc PYTHIA} )  
which mimics the effect of the CDF detector. The events are characterized
by considerable $\met$, due to the $\lspone$, more than 
one energetic jets, displaced secondary vertices due to the b-quark
jets and an isolated high
$p_T$ lepton from the top squark decay. Jet reconstruction,
tagging of b-jets and lepton ($e$, $\mu$) identification have been
done following the CDF analysis using the same parameters and
efficiencies. In particular, efficiency for tagging individual 
b-jets has been assumed to be 0.24 \cite{cdfbtag}.

The important event selection criteria following CDF are as follows:
\begin{enumerate}
  \item Considerable $\met$ due to the two $\lspone$'s  
   and a $\nu$ from the 
  decays of $\lstop \lstop^*$ : $\met \ge 25$ GeV
  \item At least 2 jets where the jets are reconstructed within 
        $|\eta| \le 2.4$ with cone algorithm 
        ($\Delta R = \sqrt{(\Delta\eta)^{2} + (\Delta\phi)^{2}} < 0.7$) : 
        $E_{T}^{jet,1} > 12$ GeV and $E_{T}^{jet,2} > 8$ GeV, where the 
        jets are ordered in descending $\met$ .
  \item At least one isolated lepton : 
            electrons with $E_{T} > 10$ GeV and $|\eta_{e}| < 1.1$ and 
            muons $p_{T} > 10$ GeV and $|\eta_{\mu}| < 0.6$ are selected.
  \item Events with opposite sign di-lepton were removed to reduce Drell-Yan 
        background.
  \item At least one secondary vertex tagged jet from one of the b-jets.
\end{enumerate}

To test the reliability of our simulation and analysis, $\ttbar$ events 
generated by
{\textsc PYTHIA} have been passed through the same simulation and
analysis chain.
 The number of $\ttbar$ events passing our selection 
is 17.38 which compares favourably with the number of  $\ttbar$ events
passing CDF selection, $\ie 17.8\pm 4.5$ ~\cite{cdfstop}.
This validates the simulation and analysis chain used for 
deriving the new limits.
Evidently the efficiency increases with increasing $\mlstop$ and for
the same $\mlstop$ increases with decreasing $\mlspone$, (see Fig.\ref{fig1}). 
Thus limits better than that obtained by the 
ALEPH Collaboration\cite{alephstop} for low $\mlspone$ is expected.

From 88 $\ipb$ data a total of 87.3$\pm$ 8.8 background events($N_b$) 
were expected from SM processes. Significant contributions come from the
$W+jets$, where W decays leptonically, $\ttbar$, $\bbbar$, 
$t \bar b$, $Z+n$-$jets$,  where 
$n\ge$ 2 and fake lepton events. Number of data events ($N_{data}$) passing
the selection is 81 (see Ref.\cite{cdfstop}).

\begin{figure}[!htb]
\vspace*{-3.5cm}
\hspace*{-3.0cm}
\mbox{\psfig{file=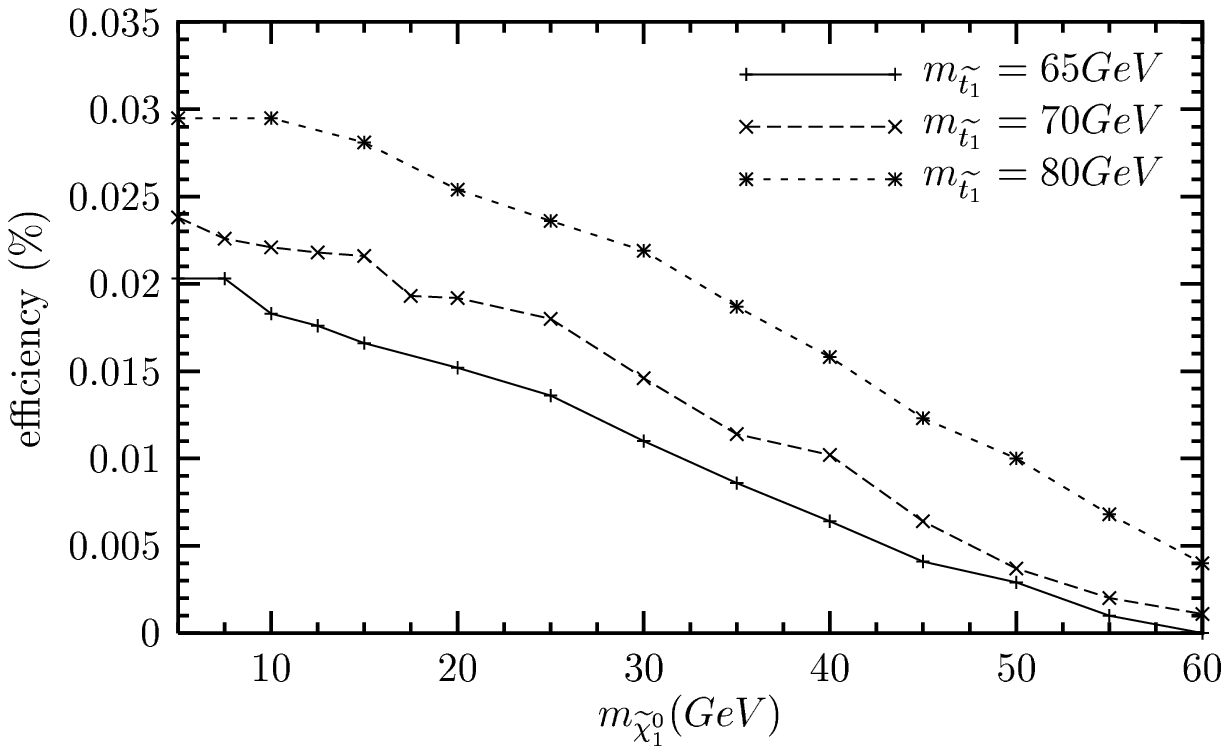,width=20cm}}
\vspace*{-16.7cm}
\caption{\small
Selection efficiency of the signal  plotted as a function of
$\mlspone$ for different values of the  top squark  mass ($\mlstop$).
}
\label{fig1}
\end{figure}

\begin{figure}[!htb]
\vspace*{-3.5cm}
\hspace*{-3.0cm}
\mbox{\psfig{file=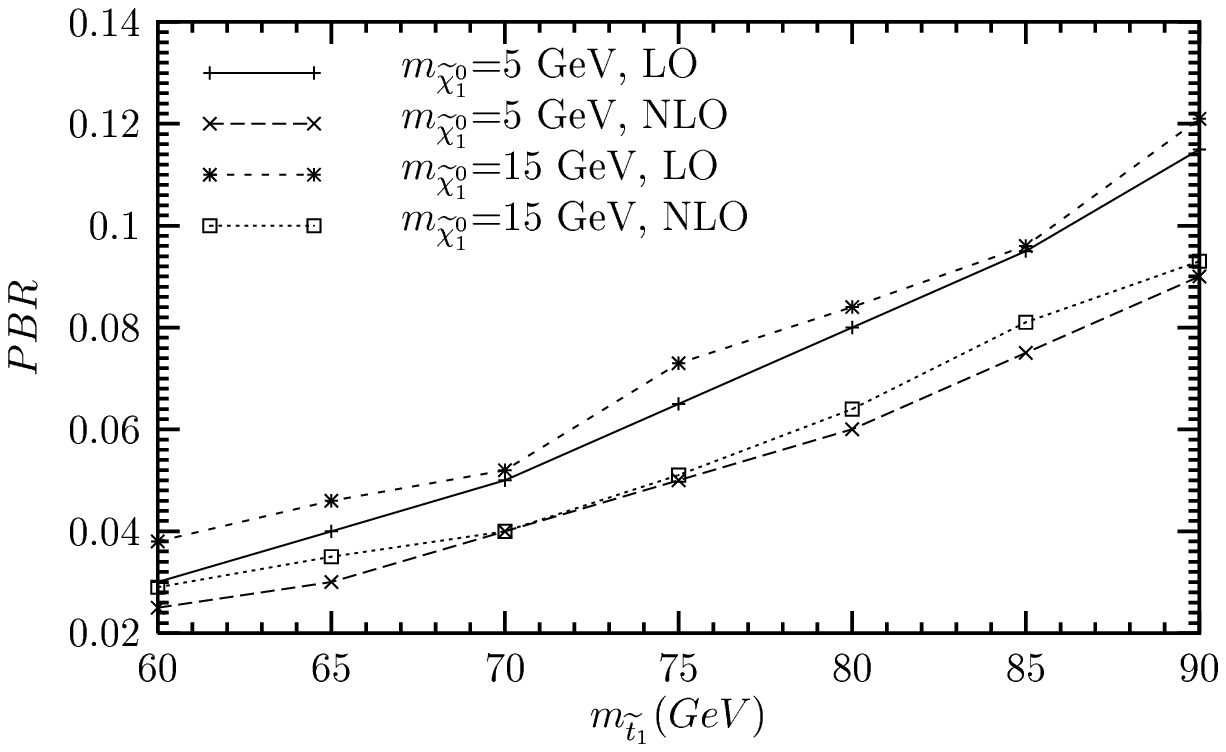,width=20cm}}
\vspace*{-16.7cm}
\caption{\small
The upper limit on the 
product of branching ratios ( PBR, see text)  as a function of 
$\mlstop$ for different values of  $\mlspone$. Limits for
both leading order(LO) 
and next to leading order(NLO) $\lstoplstop$  production cross-section
are presented.
}
\label{fig2}
\end{figure}

Since no excess over the expected SM background is observed in the data, 
95\% CL upper limits on the product of the  branching ratios
= $\branche$=$\branchmu$, where q and $q'$ correspond 
to all flavours kinematically allowed, are obtained for different values
of $\mlstop$ and $\mlspone$.  Hereafter  this  product of branching ratios
will be denoted by  PBR. For determining the 95\% CL limits, posterior 
probability distributions for each $\mlstop$ and $\mlspone$ were obtained 
using the Bayesian technique\cite{bayesian} assuming the following: error on 
the luminosity $\pm 4 \ipb $, error on the total expected number of 
SM background events $\pm 8.8$ ( taken from Ref.\cite{cdfstop} ) and $\pm 10\%$ error on the estimated acceptance 
of the signal events. We have not taken into account the error in the 
cross section due to the uncertainties in the choice of the parton
distribution functions, but we have  checked  
that even if we take into account this uncertainty no appreciable change
in the   limits  occur. For each value of the PBR 1000 Monte Carlo experiments
were performed to determine the corresponding  probability. 

\begin{figure}[!htb]
\vspace*{-3.5cm}
\hspace*{-3.0cm}
\mbox{\psfig{file=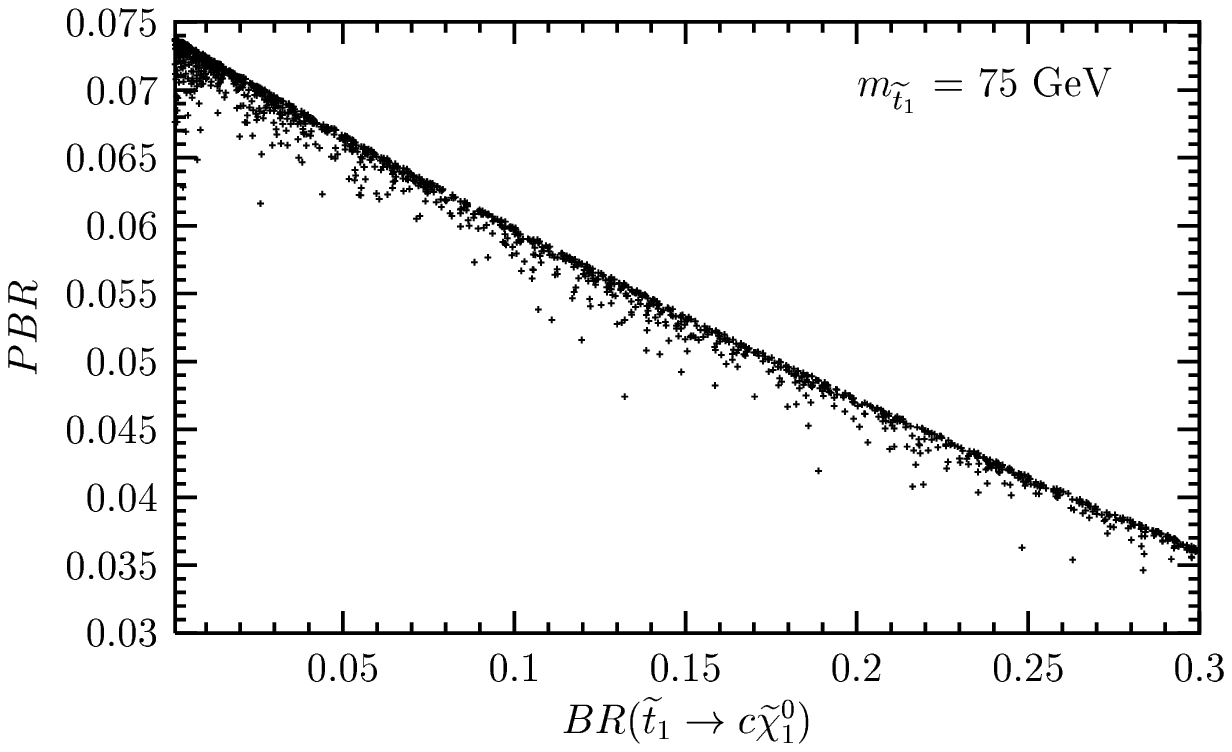,width=20cm}}
\vspace*{-16.7cm}
\caption{\small
The  PBR as a function of $\brloopdk$ for  $\mlstop$=75
GeV. The modest spread in PBR for a fixed $\brloopdk$ is due to 
the variation of the MSSM parameters (see text).
}
\label{fig3}
\end{figure}

Upper limits on the PBR have been calculated using  $\sigma(\ppbar\ra\lstoplstop)$
both in the LO and NLO approximation. 
These limits  are shown  in Fig.\ref{fig2}. For any given 
combination of $\mlstop$ and $\mlspone$ the limits
obtained by using the relatively low LO production cross section are weaker
as expected. We shall follow the prescription of
Ref.\cite{nlostop} and  estimate the possible impact of the NLO
cross section on our limits  by using  the LO cross
sections scaled by a factor of 1.3 as an input to the limit calculation.
The resulting stronger limits are  presented  in
Fig.\ref{fig2}. We have concentrated on low values of $\mlspone$ because
in this region of the parameter space our analyses lead to limits better 
than those obtained by ALEPH\cite{alephstop}. 

The regions of the MSSM parameter space, where the theoretical
expectation of the PBR is above  the $95\%$ CL upper limit are
excluded by this analysis.  We shall now show that a
significant region of the MSSM  parameter space which was not excluded 
by the previous analyses are excluded in a fairly model independent way.

\begin{figure}[!htb]
\vspace*{-3.5cm}
\hspace*{-3.0cm}
\mbox{\psfig{file=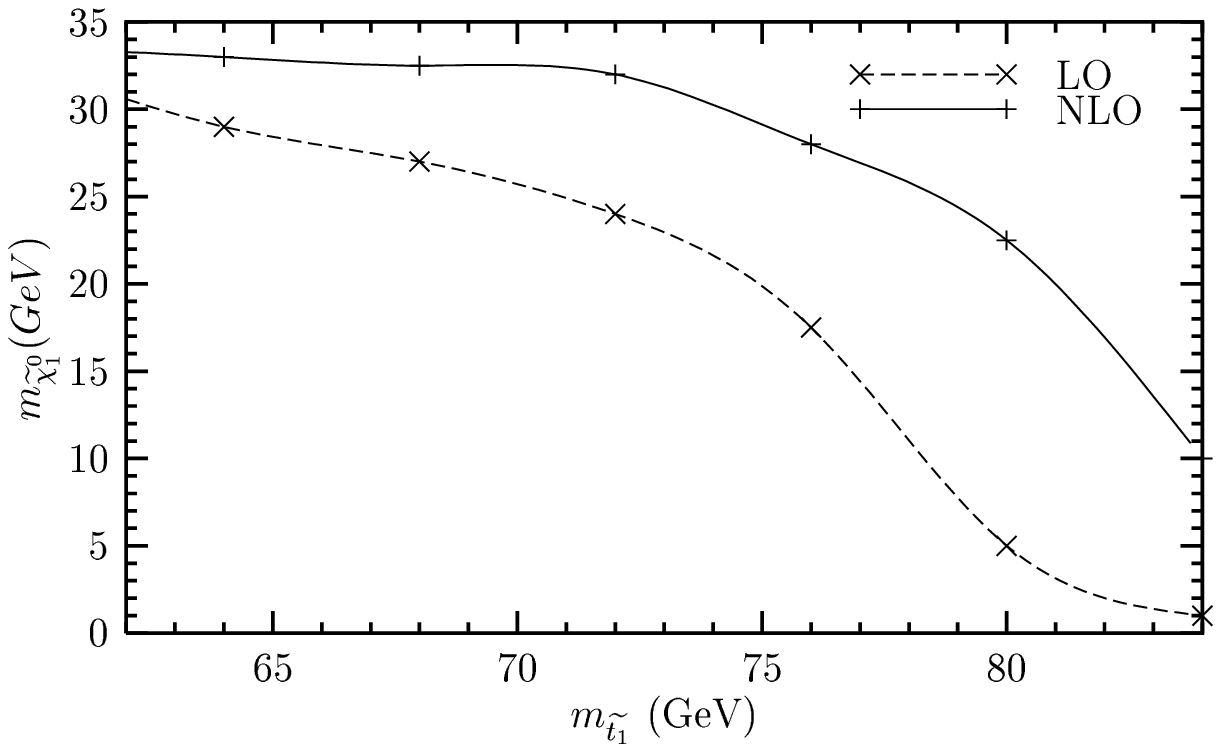,width=20cm}}
\vspace*{-16.7cm}
\caption{\small
The excluded  region in the $\mlstop -\mlspone$ plane from  Tevatron Run-I
assuming the $\brloopdk$ to be  negligible. The regions below the dotted
(solid) lines are excluded using the LO (NLO) production cross sections. 
}
\label{fig4}
\end{figure}

%%%
\begin{figure}[!htb]
\vspace*{-3.5cm}
\hspace*{-3.0cm}
\mbox{\psfig{file=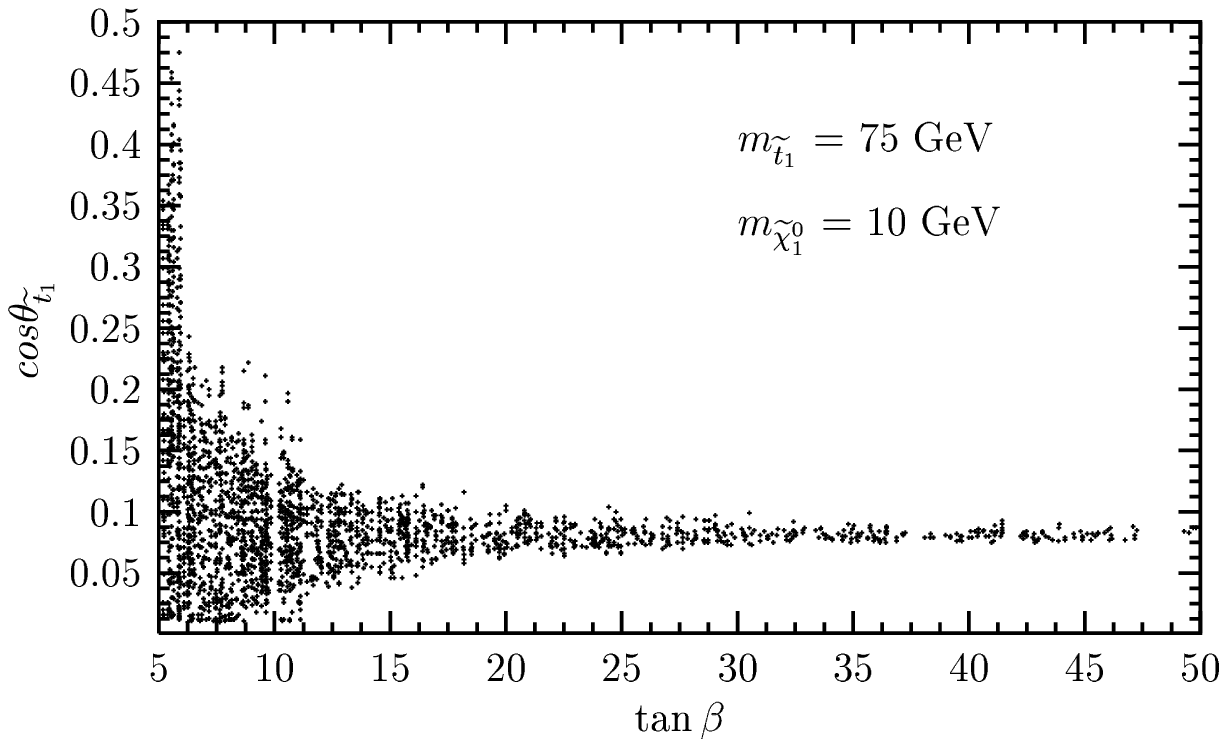,width=20cm}}
\vspace*{-16.7cm}
\caption{\small
The excluded regions in the $\tb -\cost$ plane for  
$\mlstop$=75 GeV and $\mlspone$=10 GeV. 
The  dots  represent excluded points  where the PBR exceeds its
upper limit for  this 
combination of $\mlstop$ and  $\mlspone$.
 $M_{1}, M_{2}$ and  $\mu$ are varied
such that  the  $\mlspone$ is fixed. The other MSSM parameters, which
affects the result mildly (see text)
are  $\mslepl$=$\mslepr$= 300 GeV, 
$\msql$=$\msqr$=500 GeV, $\abot$=300 GeV , $\atau$=200 GeV, $M_{A}$=300 GeV.
}
\label{fig5}
\end{figure}

In Fig.\ref{fig3}, we plotted the PBR as a function of the loop decay
BR for $\mlstop$ =75 GeV. The two BR's in the product 
are calculated  by randomly varying
all other MSSM parameters, taking into account the  direct limits from 
LEP and Tevatron\cite{lepbound,tevbound,lepboundstop}. In particular 
the following ranges have been considered: the common slepton mass
$\mslepl$=$\mslepr$= [120 - 500] GeV, the common mass for the squarks 
$\msql$=$\msqr$=[300 - 800] GeV,
$\cost$=[0.01 - 0.90 ] ( $\mix$=[89.43$^{\circ}$ - 25.84$^{\circ}$] ) , 
the trilinear soft breaking term in the b sector
$\abot$=[150 - 750 ] GeV , the trilinear soft breaking term in the $\tau$ 
sector $\atau$=[150 - 350 ] GeV, the U(1) gaugino mass  $M_{1}$=[5 -
50] GeV, the SU(2) gaugino mass
$M_{2}$=[110 - 300] GeV, the higgsino mass parameter
$\mu$=[50 - 500 ] GeV, the ratio of the vacuum expectation values of  
the two neutral Higgs fields $\tb$=[5 - 50 ] and the pseudoscalar Higgs
boson mass $M_{A}$=[300 - 900] GeV. It should be noted that we have not
invoked the model dependent assumption of gaugino mass unification.
On the other hand the common mass for the first two generations
of squarks as suggested by the absence of flavour changing neutral
currents has been used.  
We have also checked that the PBR remain  almost unchanged
if we consider the range $\mu$=[(-500) - (-50)] GeV.  Hence, the figure is
drawn only for positive $\mu$.

A cursory look into the Feynman diagrams \cite{boehm}
for each of the 4-body decay
mode of $\lstop$ would suggest that the theoretical prediction 
of the above product depends on many free parameters.
An important result of this letter is to establish that inspite of this 
apparent model dependence, a fairly model independent approach for 
extracting the limits is possible.
For a fixed BR($\loopdk$), 
the PBR lies in a rather narrow range even if all model  parameters
are arbitrarily varied. This is not difficult to understand.
For top squark masses sensitive to the data we are examining  and
chargino
and slepton masses above the current LEP limits, both the semileptonic
and the hadronic 4-body BR's  follow the corresponding $W^*$ BR 
in most cases. For a given loop decay BR's, the overall 4-body BR's  
is fixed. Now the PBR lies in a narrow range even if the
MSSM parameters are widely varied subject to the above constraints. We have 
found the same behaviour  of the PBR for other value of $\mlstop$ 
relevant for our analysis ( $65.0 \lapp \mlstop \lapp 85.0 GeV $).

Some numerical examples are given below. When the loop decay BR is 
negligible ($\lapp 0.5 \% $),
the theoretical PBR lies between 0.069 - 0.073. This range reflects the
uncertainty in the  
MSSM parameters, $\cost$ being the most important one 
among them. In this 
particular case if $0.01 < \cost < 0.18$  ($ 89.43^{\circ} > \mix > 
79.63^{\circ}$ ), the BR  of the loop decay is negligible.
For LSP mass 5 (15) GeV the limit is $\mlstop >$ 76.5 - 77.5 GeV 
( 74.0 - 75.0 GeV ) if the limiting PBR  ( see Fig.\ref{fig2})  
corresponding to the LO production cross section  is used. 
For the purpose of quoting limits we shall use the central value
 of each range. 

Using the NLO cross section as discussed above  the corresponding limits
become stronger: 83.0 - 84.2 GeV ( 81.5 - 82.5 GeV) for $\mlspone$ = 5(15) GeV.
Thus most of the narrow region of the parameter space corresponding to 
large $\mix$ and low $\mlspone$ allowed by the ALEPH analysis
( see Fig.3(a) of Ref.\cite{alephstop}), is
disallowed by Run-I data if the NLO cross section is used.

If on the other hand the loop BR's is fixed at 20 \%,  
for LSP mass 5 (15) GeV, the limit is $\mlstop >$
67.5  - 68.5  GeV ( 64.2 - 66.1  GeV ) using the LO cross section.
Stronger limits 72.5 - 73.5  GeV ( 72.2 - 73.2 GeV ) emerge  
corresponding to the NLO cross section. This should be compared 
with Fig.4(b) of Ref.\cite{alephstop}.

Assuming that the loop decay BR's is negligible and fixing the PBR at
0.073, a number motivated by Fig.\ref{fig3}, we present the constraints
in the $\mlstop - \mlspone$ plane in  Fig.\ref{fig4}. This may be compared 
with the limits in Fig.3(a) of ~\cite{alephstop}. Although the improvement 
using
the leading order cross section is rather modest, the NLO cross
section leads to significant improvement in the large $\mix$ and
small $\mlspone$ scenario. Our results, therefore, nicely 
complements the ALEPH limits.

Some comments on the importance of the parameter $\cost$ are now in order.
It has already been noted in Ref.\cite{boehm} that the parameter
$\epsilon$, as defined in \cite{hikasa}, plays a crucial role in the loop
decay. By adjusting various SUSY parameters the magnitude of
$\epsilon$ can be suitably reduced
leading to a vanishingly small loop decay BR. 
In some regions of the parameter space this may require a fair amount
of fine-tuning. This is illustrated in Fig.\ref{fig5} in the 
$\cost-\tb$ plane for $\mlstop$= 75 GeV and $\mlspone$=10 GeV, 
where the dots correspond to the 
PBR greater than or equal to its limiting value.   The parameters 
$M_1, M_2$ and  $\mu$ have been  varied such that the $\mlspone$ is fixed. 
The choice of the other MSSM parameters are
given in the figure caption. It is seen that for large $\tb$
the PBR is sensitive to the data for a very narrow range
of $\cost$ values,  where as for small $\tb$ this happens for
a much larger range of $\cost$. The dominance of the 4-body
is, therefore, more probable at small $\tb$. For example in 
Fig.\ref{fig5}, the PBR is greater than or equal to its limiting value 
for  $0.07 < \cost < 0.1( 85.98^{\circ} > \mix > 
84.26^{\circ} )$  and $\tb$=45. Even if all MSSM parameters  are randomly
varied keeping $\mlstop$ and $\tb$ fixed, the above range
marginally changes 
to   $0.01 < \cost < 0.1( 89.43^{\circ}  > \mix > 84.26^{\circ} )$. 
On the other hand for   $\tb$=5  the range for the above set
of MSSM  parameters 
is $0.01 < \cost < 0.35( 89.43^{\circ}  > \mix >
69.51 ^{\circ}) $, see Fig.\ref{fig5} . These features have been
noted for all $\mlstop$ sensitive to the  data we are using.

%%%%\section*{4.~ Conclusions}

 This letter  sketches a fairly model independent strategy for top
squark search including its 4-body decays. This approach can be easily
extended  to anayles using Run-II data.
  There are reasons to be optimistic about the 
prospect of $\lstop$ search via the 4-body decay channels at Tevatron Run-II.
Firstly the  $\lstoplstop$ production cross-section at $\rts$=2 TeV
will be  slightly higher and the total integrated 
luminosity much larger. Moreover  kinematical cuts  can be specially 
designed for the 4-body decay channel. For example, $jets + \met$
data at a relatively low $\met$ will improve the search 
prospect via the hadronic decay mode of both the top squarks, which has the 
largest BR's. Instead of using a very stiff  $\met$ cut, the background 
can be suppressed by efficient b-tagging and utilizing the large number of jets
in the signal. With good $\tau$ detection efficiency  4-body final states 
with  $\tau$-leptons may also lead to further improvements.

\noindent
{\bf Acknowledgements :} AD acknowledges financial support from 
BRNS(INDIA) under the project number 2000/37/10/BRNS.
SPD acknowledges the grant of a senior fellowship by CSIR, India. 
the authors thank  N.K.Mondal, S.Chakrabarti and S.Jain 
for helpful discussions.

\end{document}